\documentclass[preprint2]{aastex631}
\submitjournal{ApJ}

\shorttitle{A UFO Seen Edge-On in Mrk 34}
\shortauthors{Maksym et al.}
\usepackage{hyperref,breakurl}
\def\UrlBreaks{\do\/\do-}
\usepackage{ifthen,psfig,graphicx}
\usepackage{epstopdf,dblfloatfix}
\usepackage{soul,color,amsmath}
\usepackage{ifthen,psfig,graphicx}
\usepackage{epstopdf,overpic}
\usepackage{soul,color,amsmath}
\usepackage{tikz}
\usepackage{placeins}
\usepackage[T1]{fontenc}
\def\mathnew{\mathsurround=0pt}
\def\simov#1#2{\lower 2.5pt\vbox{\baselineskip0pt \lineskip-.5pt
\ialign{$\mathnew#1\hfil##\hfil$\crcr#2\crcr\sim\crcr}}}
\def\simless{\mathrel{\mathpalette\simov <}}
\def\simgreat{\mathrel{\mathpalette\simov >}}
\newcommand{\MeV}{Me\kern-0.11em V}
\newcommand{\keV}{ke\kern-0.11em V}

\newcommand{\z}{{\it z\/}}

\newcommand{\white}[1]{\white{white}{#1}}
\makeatletter
\newcommand\iont[2]{{#1$\;${\small\expandafter\@slowromancap\romannumeral #2@\relax}}}
\makeatother
\makeatletter
\newcommand{\raisemath}[1]{\mathpalette{\raisem@th{#1}}}
\newcommand{\raisem@th}[3]{\raisebox{#1}{$#2#3$}}
\makeatother

\begin{document}

\title{A UFO Seen Edge-On: \\ Resolving Ultrafast Outflow Emission on $\sim$200-pc Scales with {\it Chandra}\\  in the Active Nucleus of Mrk 34 }

\correspondingauthor{W. Peter Maksym; @StellarBones}
\email{walter.maksym@cfa.harvard.edu}

\author[0000-0002-2203-7889]{W. Peter Maksym}
\affiliation{Center for Astrophysics \textbar\ Harvard  \& Smithsonian, 60 Garden St., Cambridge, MA 02138, USA}

\author[0000-0001-5060-1398]{Martin Elvis}
\affiliation{Center for Astrophysics \textbar\ Harvard  \& Smithsonian, 60 Garden St., Cambridge, MA 02138, USA}

\author[0000-0002-3554-3318]{Giuseppina Fabbiano}
\affiliation{Center for Astrophysics \textbar\ Harvard  \& Smithsonian, 60 Garden St., Cambridge, MA 02138, USA}

\author[0000-0001-8112-3464]{Anna Trindade-Falc\~ao}
\affiliation{Center for Astrophysics \textbar\ Harvard  \& Smithsonian, 60 Garden St., Cambridge, MA 02138, USA}

\author[0000-0003-4073-8977]{Steven B. Kraemer}
\affiliation{Institute for Astrophysics and Computational Sciences, Department of Physics, The Catholic University of America, Washington, DC 20064, USA}

\author[0000-0002-3365-8875]{Travis C. Fischer}
\affiliation{Space Telescope Science Institute,
3700 San Martin Drive, Baltimore, MD 21218, USA}

\author{D. Michael Crenshaw}
\affiliation{Department of Physics and Astronomy, Georgia State University, Astronomy Offices, 25 Park Place, Suite 600, Atlanta, GA 30303, USA}

\author[0000-0003-1772-0023]{Thaisa Storchi-Bergmann}
\affiliation{Departamento de Astronomia, Universidade Federal do Rio Grande do Sul, IF, CP 15051, 91501-970 Porto Alegre, RS, Brazil}



\begin{abstract}

We present {\it Chandra} ACIS imaging spectroscopy of the nucleus of the Seyfert 2 Galaxy Mrk 34.  We identify spatially and spectrally resolved features in the band that includes Fe K$\alpha$, \ion{Fe}{25} and \ion{Fe}{26}.  These features indicate high-velocity ($\simgreat15,000\,\rm{km\,s}^{-1}$ line-of-sight) material spanning $\sim0\farcs5$, within $\sim200$\,pc of the nucleus.  This outflow could have deprojected velocities $\sim12-28\times$ greater than the [\ion{O}{3}] emitting outflows, and could potentially dominate the kinetic power in the outflow.  This emission may point to the origins of the optical and X-ray winds observed at larger radii, and could indicate a link between ultra-fast outflows and AGN feedback on $\simgreat$kpc scales.  

\end{abstract}


\section{Introduction} \label{sec:intro}

Active galactic nuclei (AGN) are observed to generate jets and winds, which are thought to play critical roles in galaxy evolution in the form of `feedback', which may stimulate or suppress star formation via the expulsion of star-forming gas from the host galaxy, by regulating the temperature and density of the remaining ISM, and by affecting the rate at which the nuclear black hole accretes matter.  

AGN feedback is likely a complex, multi-stage process \citep{HE10,GS17,Harrison18}.  Ultra-fast outflows seen in absorption \citep[UFOs,][]{Laha21} reach $\simgreat0.2c$, which is clearly capable of expelling ISM.  But even the closest UFOs are spatially unresolved and inferences that such velocities are sustained beyond $\sim$pc scales therefore require indirect constraints.  Large-scale outflows in the host ISM may reach $\simgreat$kpc scales but rarely exceed $\sim2000\,{\rm km\,s}^{-1}$ in spatially resolved optically-emitting or molecular gas \citep[e.g.][]{Fischer13}.  {In the case of [\ion{O}{3}] kinematics, only velocities in excess of typical rotational velocities ($\sim100-300\,\rm{km\,s}^{-1}$) can be attributed to outflows capable of producing effective AGN feedback \citep{Fischer17}}.  Understanding the transition between these two regimes is critical to modeling the physics of AGN feedback, but this interface regime is particularly challenging to observe due to the the low emissivity of hot plasmas and the instrumental limits of modern X-ray telescopes.    

A growing number of observations indicate evidence for such an interface region at $\sim100$\,pc scales.  BALQSO spectra indicating UFO-like velocities may be common at $\simgreat100$\,pc scales \citep{Arav18}. Possible evidence for multiphase gas {\it entrained} in ultrafast outflows (E-UFOs) has also been seen in unresolved spectra at $\simgreat100$\,pc scales \citep{Serafinelli19}.

With the availability of increasingly deep {\it Chandra} observations, the Fe\,K$\alpha$ complex has recently become a promising tool for studying spatially resolved AGN-host interactions on sub-kpc scales.  This complex includes neutral Fe\,K$\alpha$ (6.4\,keV rest energy) as well as helium-like \ion{Fe}{25} triplet at $\sim6.7$\,keV and the 7.0\,keV Ly$\alpha$ of hydrogenic \ion{Fe}{26}.  A growing number of observations of local Seyfert 2 galaxies like NGC 6240, ESO 428-G014, IC 5063, NGC 3393 and others have shown extended hard (3-8 keV) emission, extending to $\sim$kpc scales in some cases \citep{Fabbiano17,Fabbiano18a,Fabbiano18b,Fabbiano19, Jones20, Jones21, Ma20, Maksym17, Maksym19, Travascio21,Wang14}, as well as reflected Fe\,K$\alpha$ \citep{Fabbiano17,Fabbiano19,Jones20,Travascio21} and \ion{Fe}{25} excited via collisions between AGN outflows and molecular clouds in the ISM \citep{Travascio21}.

Mrk 34 ($z=0.05080$; Hubble distance $D_{H}= 209$\,Mpc; 1\arcsec = 1014 pc for $H_0=71\,\rm{km\,s}^{-1}$, \citealt{Revalski18}) is known to have powerful outflows associated with spatially extended radio emission and narrow line emission in the optical and X-rays \citep{Falcke98,Revalski18,Revalski19,TF21,BRA22}.  A radial increase in outflow gas mass in Mrk 34, with increasing $r$ towards $r\sim500$\,pc, leads \cite{Revalski18,Revalski19} to suggest that [\ion{O}{3}] gas is accelerated {\it in situ}.  But models by \cite{TF21} indicate that the [\ion{O}{3}] gas may originate from a lower-ionization, higher-density state, and that this process may be facilitated by an X-ray wind which can be more kinematically powerful than the optical gas by a factor of $\sim50$.

Mrk 34 is highly obscured (log\,$[n_{\rm H}/{\rm cm}^{-2}]\simgreat24.3$; \citealt{Gandhi14,Zhao20}), and is therefore another excellent candidate for similar spatial mapping of extended hard emission.  In this paper, we examine spatially resolved emission in {\it Chandra} observations of Mrk 34 in the 5.5-7.0\,keV band.  We find evidence for high-velocity iron outflows which may provide a link between ultrafast outflows seen in absorption and kpc-scale winds seen in emission.

\section{Observations and Processing}

Mrk 34 was observed by {\it Chandra} with ACIS on 2017 January 30 to spatially resolve X-ray emission from termination shocks in two of the most powerful kinematic outflows with angular scales accessible to {\it Chandra} (Obsid 18121, PI: Elvis).  Detailed spatial and spectroscopic analysis of the other program target, Mrk 78, was published by \citep{Fornasini22} and is similarly {\it in preparation} for the Mrk 34 observations (Maksym et al).  Data were reduced using standard {\it Chandra} methods and version 4.13 of the {\tt CIAO} package \citep{CIAO}, with CALDB version 4.9.3.  We reprocessed the data using {\tt chandra\_repro} and found no significant background flares.  Useful exposure time totaled 99.1\,ks.  We used {\tt wavdetect} to identify matches between field X-ray sources and SDSS \citep{SDSS}, and cross-registered the astrometry with {\tt wcs\_match} and {\tt wcs\_update}.{We excluded matches with residuals $>2\arcsec$, leaving 3 sources with positional uncertainties of $\sim0\farcs1$ each}.

Mrk 34 was previously observed by the VLA on 1991 Sep. 1 in the X-band (8.44 GHz) as part of program AW0278.  We used the processed NRAO archival image which has a beam size of $0\farcs 2713$ and rms sensitivity of 26 $\mu$Jy.

\section{Data Reduction and Results}

\subsection{Nuclear Spectroscopy}

%\begin{figure}[th!]
\begin{figure}
\vspace{0.1in}
\noindent
\centering
\includegraphics[width=0.47\textwidth]{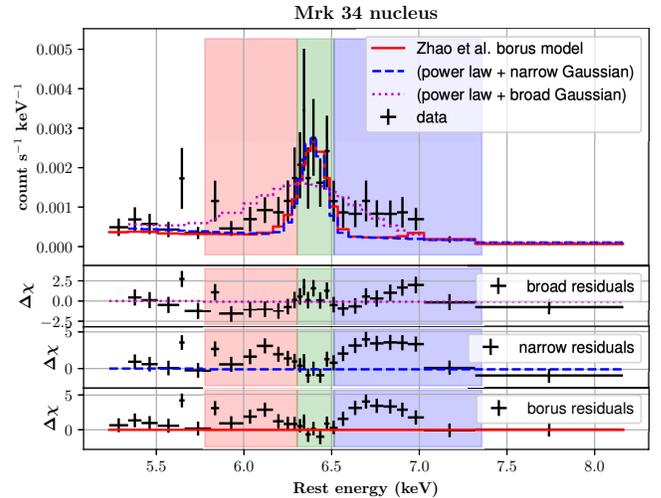}
%\begin{tikzpicture}
%\node[anchor=south west,inner sep=0] at (0,0) {\includegraphics[width=0.47\textwidth]{f1.eps}};
%\filldraw[color=red, very thin, fill=red,opacity=0.25](2.62,0.62) rectangle (3.75,6.47);
%\filldraw[color=green, very thin, fill=green,opacity=0.25](3.75,0.62) rectangle (4.23,6.47);
%\filldraw[color=blue, very thin, fill=blue,opacity=0.25](4.24,0.62) rectangle (6.05,6.47);
%\end{tikzpicture}
\caption{
{\it Chandra} 5-8 keV spectrum for $r<2\arcsec$ of the Mrk 34 nucleus.  The red solid model uses the borus and absorbed power law components used by \citep[][red line]{Zhao20}, normalized to the Fe K$\alpha$ peak between 6.3\,keV and 6.5\,keV (rest frame; green shading) and with no thermal plasma component.  {Magenta dots and blue dashes indicate a simpler absorbed power law plus Gaussian, with the Gaussian width free to vary (magenta dots) or assumed narrow and unresolved (blue dashes).  All models produce systematic residuals. Residuals compared to \cite{Zhao20} and the narrow Gaussian are significant compared} to the red (5.8-6.3\,keV, rest) and blue (6.5-7.4\,keV, rest) of Fe\,K$\alpha$, and includes the rest energies of \ion{Fe}{25} and \ion{Fe}{26}.
} 
\label{fig:spec}
\end{figure}

In order to identify useful line emission features for sub-arcsecond imaging, we {initially} assumed the active galactic nucleus to be located at the centroid of the 6-7\,keV band {for the purpose of spectroscopy}.  We extracted a spectrum of the nucleus for a $r=2\arcsec$ circle at this position (J2000 $[\alpha,\delta]=[10h34m08.59s,+60\degr01\arcmin52\farcs1]$), which is comparable to the $\sim90\%$ encircled energy radius for this band.  The spectrum has 5 photons per energy bin, which when fitted with {\it lstat} \citep{Loredo92} is a useful compromise between the energy resolution of unbinned {\it cstat} fitting \citep{Cash79} and the ability to subtract an unmodeled background with $\chi^2$ fitting.

We fit this spectrum with {\tt XSpec}, excluding the soft emission below 3\,keV (observer) which is typically dominated by ionized reflection or emission from collisionally excited plasma on scales reaching the extended narrow line region (ENLR $\simgreat$\,pc scales; see e.g. \citealt{Maksym17,Maksym19}  {and for Mrk 34, \citealt{TF21} and Maksym et al. {\it in prep}}). 

We can model the continuum with a simple absorbed power law model fit to 3-8\,keV (observer) which excludes the Fe\,K$\alpha$ complex by ignoring data points between 5.5-7.0\,keV (observer).  We then model an expected neutral Fe\,K$\alpha$ contribution to the observed excess emission by adding {a ($\sigma=0.01$\,keV) Gaussian at $E=6.39$\,keV (rest).}

The best fit for this model produces an effective 5.5-7.0\,keV (observer) $\Delta\chi^2/\rm{DoF}=28.14/28$ and Gaussian width $\sigma=(13\pm2)\times10^3\,\rm{km\,s}^{-1}$.  This is an order of magnitude larger than velocities characteristic of the torus and 

\clearpage
\makeatletter\onecolumngrid@push\makeatother
\begin{figure*}[p]
%\vspace{0.1in}
\noindent
\centering
\begin{overpic}[scale=0.29]{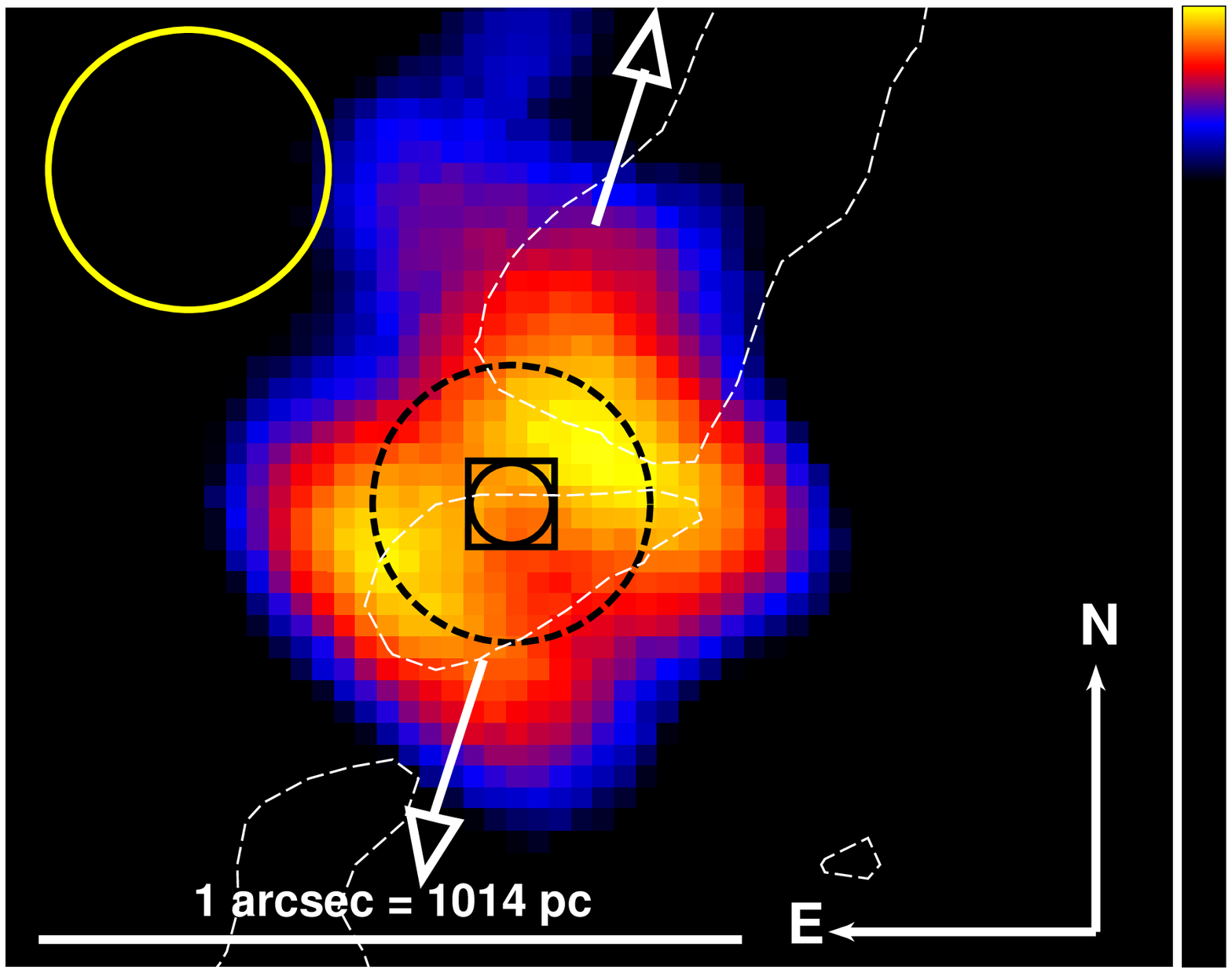}
	\put(55,65){\color{white}\bf 3.2-5.8\,keV}
\end{overpic}%
\begin{overpic}[scale=0.29]{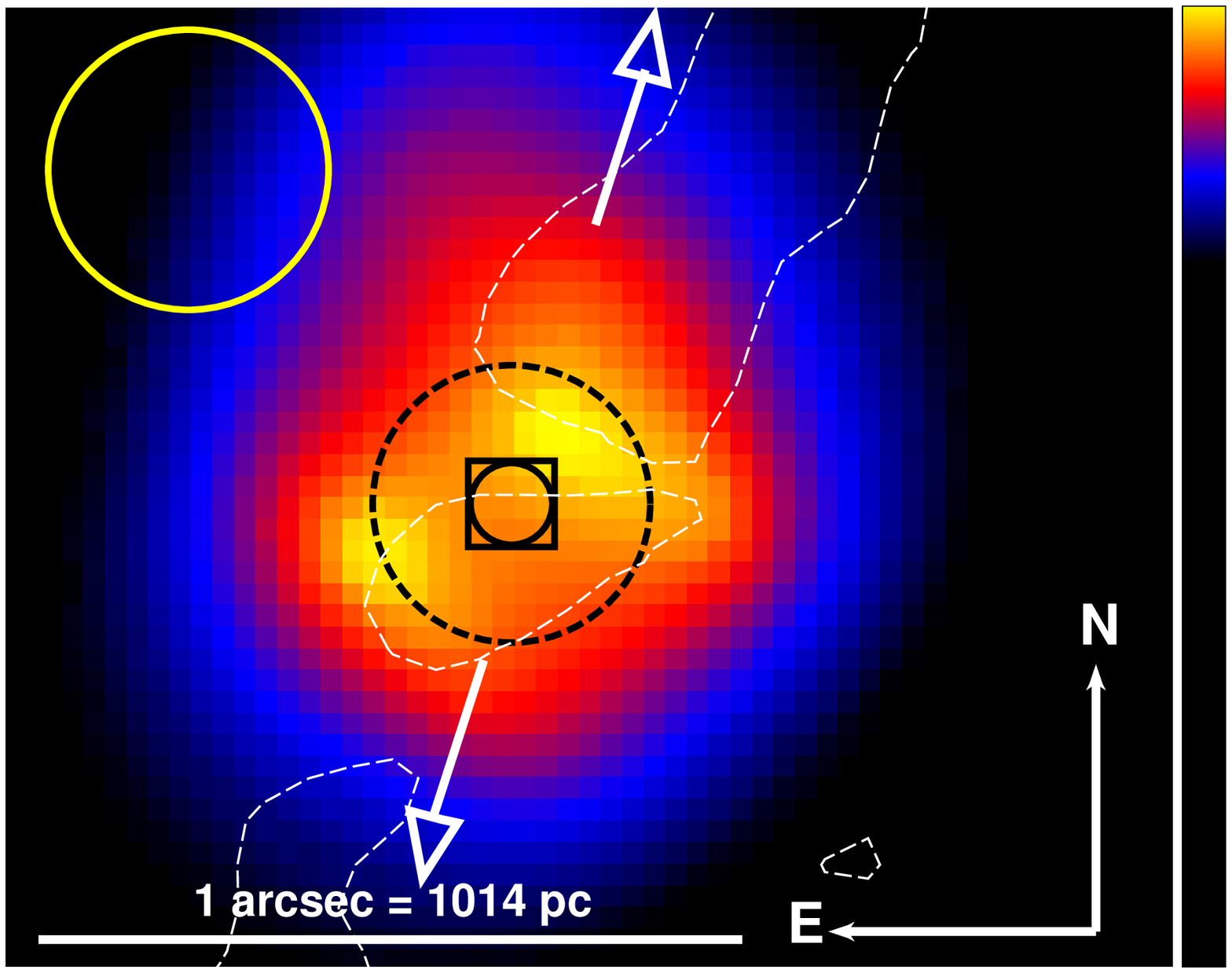}
	\put(55,65){\color{white}\bf 3.2-5.8\,keV}
\end{overpic}%
\includegraphics[width=0.29\textwidth]{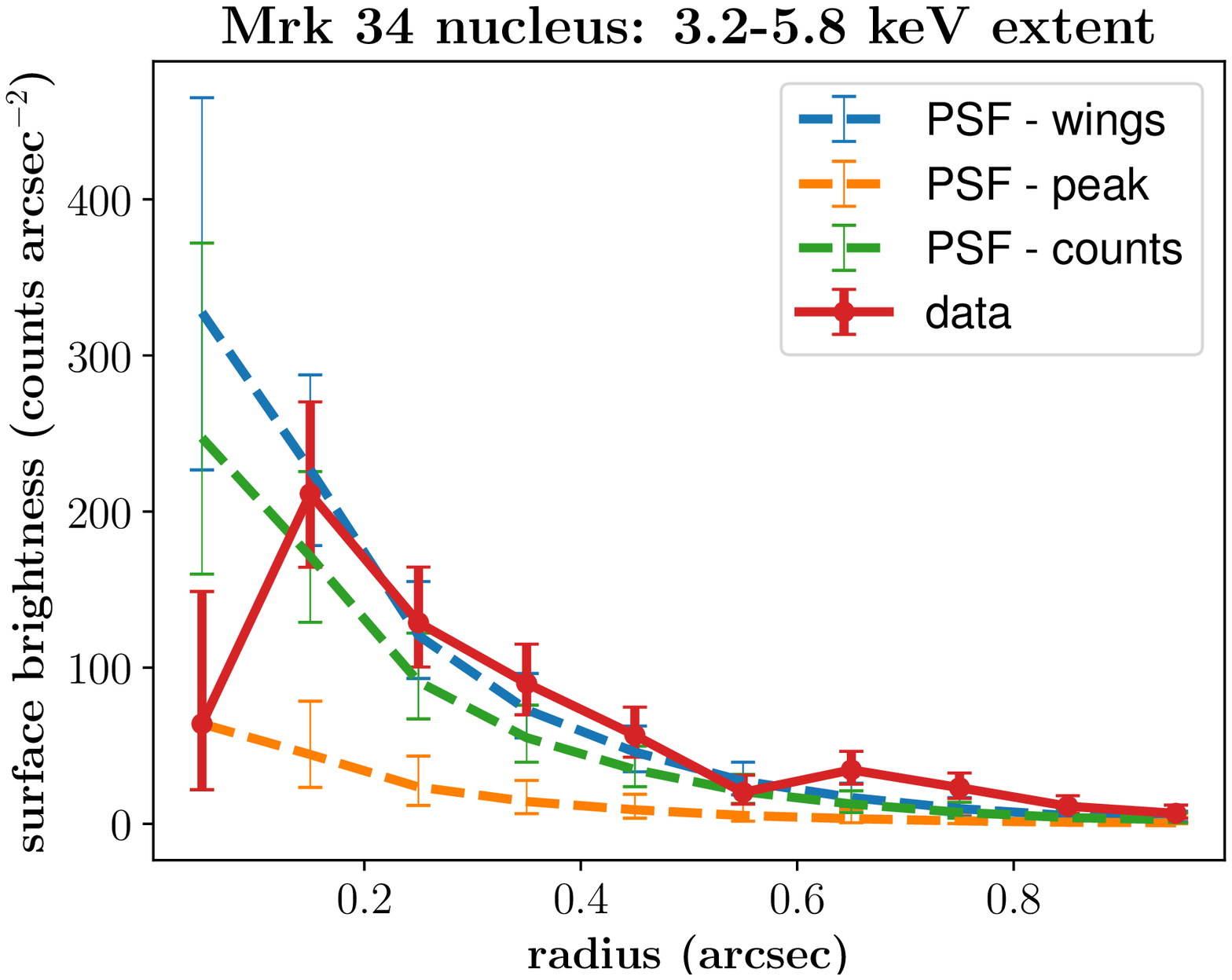}\\
\begin{overpic}[scale=0.29]{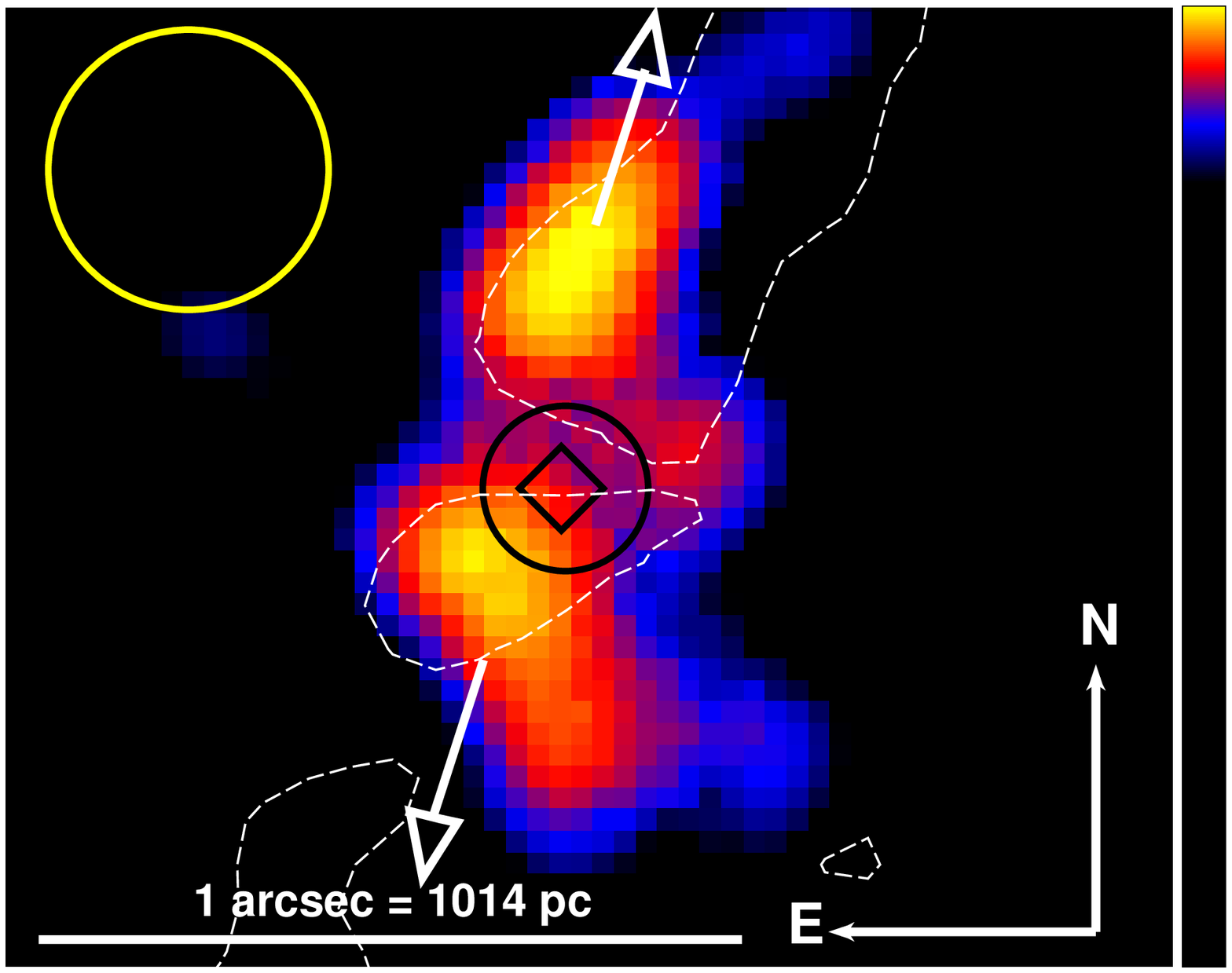}
	\put(55,65){\color{white}\bf 5.8-6.3\,keV}
\end{overpic}%
\begin{overpic}[scale=0.29]{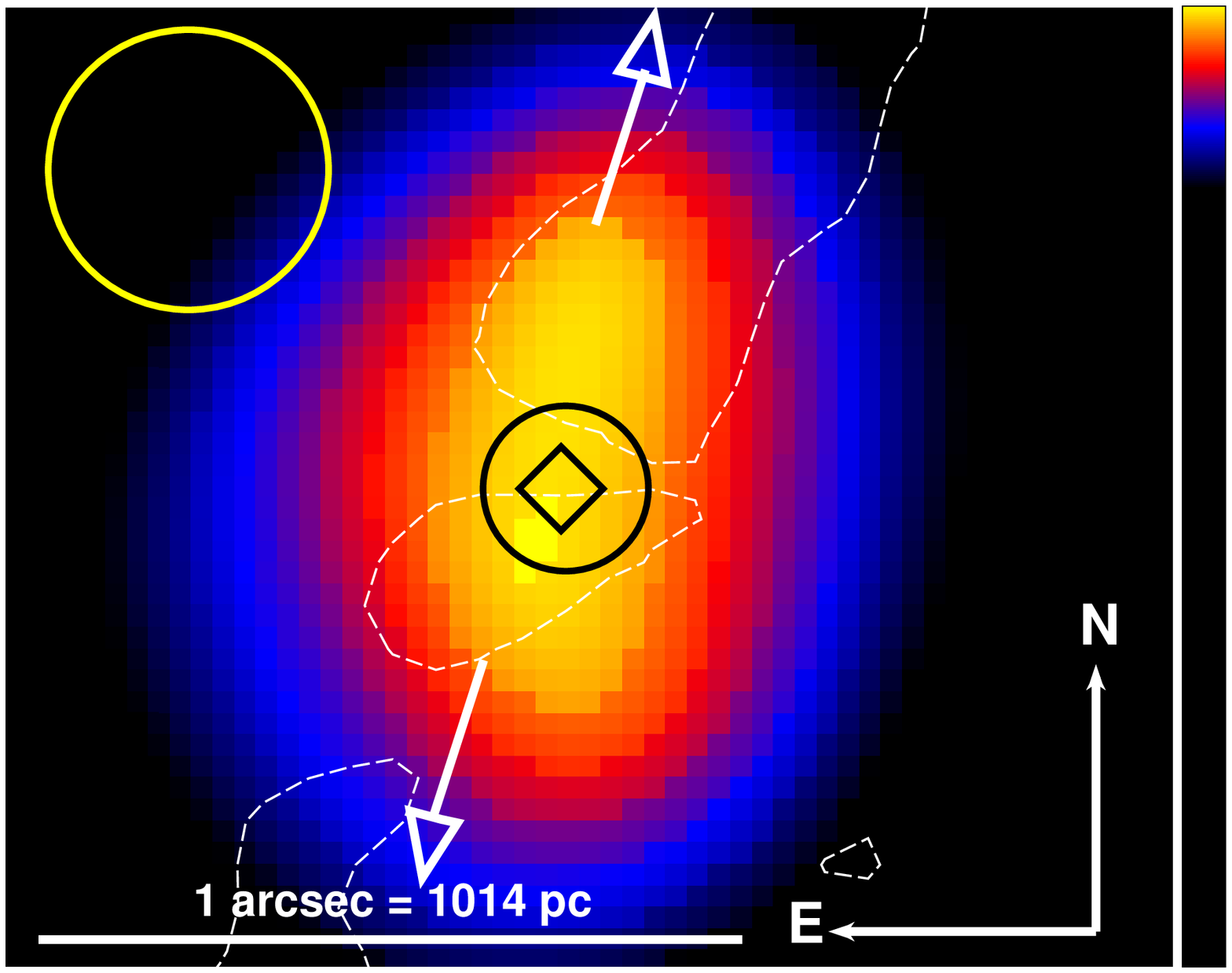}
	\put(55,65){\color{white}\bf 5.8-6.3\,keV}
\end{overpic}%
\includegraphics[width=0.29\textwidth]{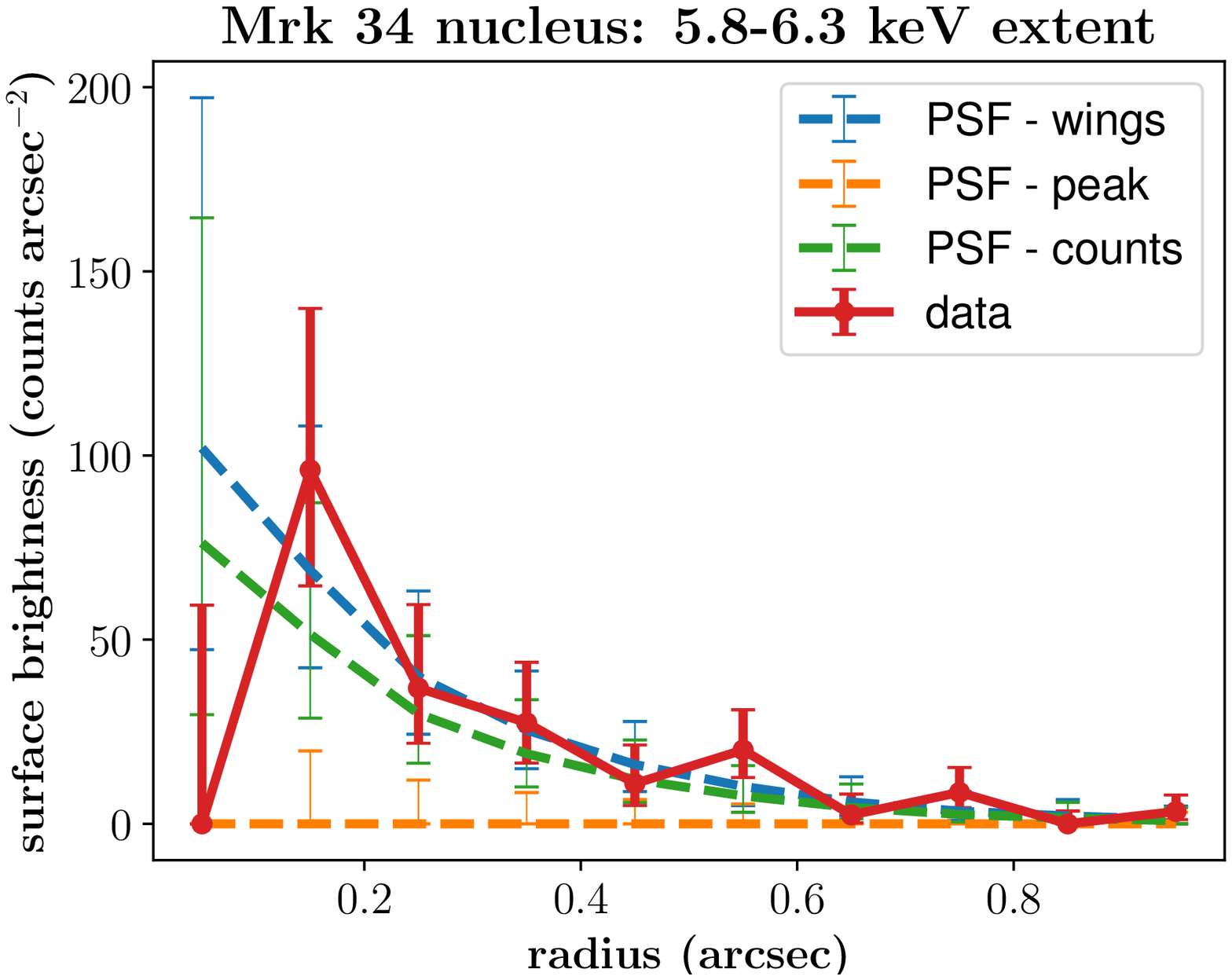}\\
\begin{overpic}[scale=0.29]{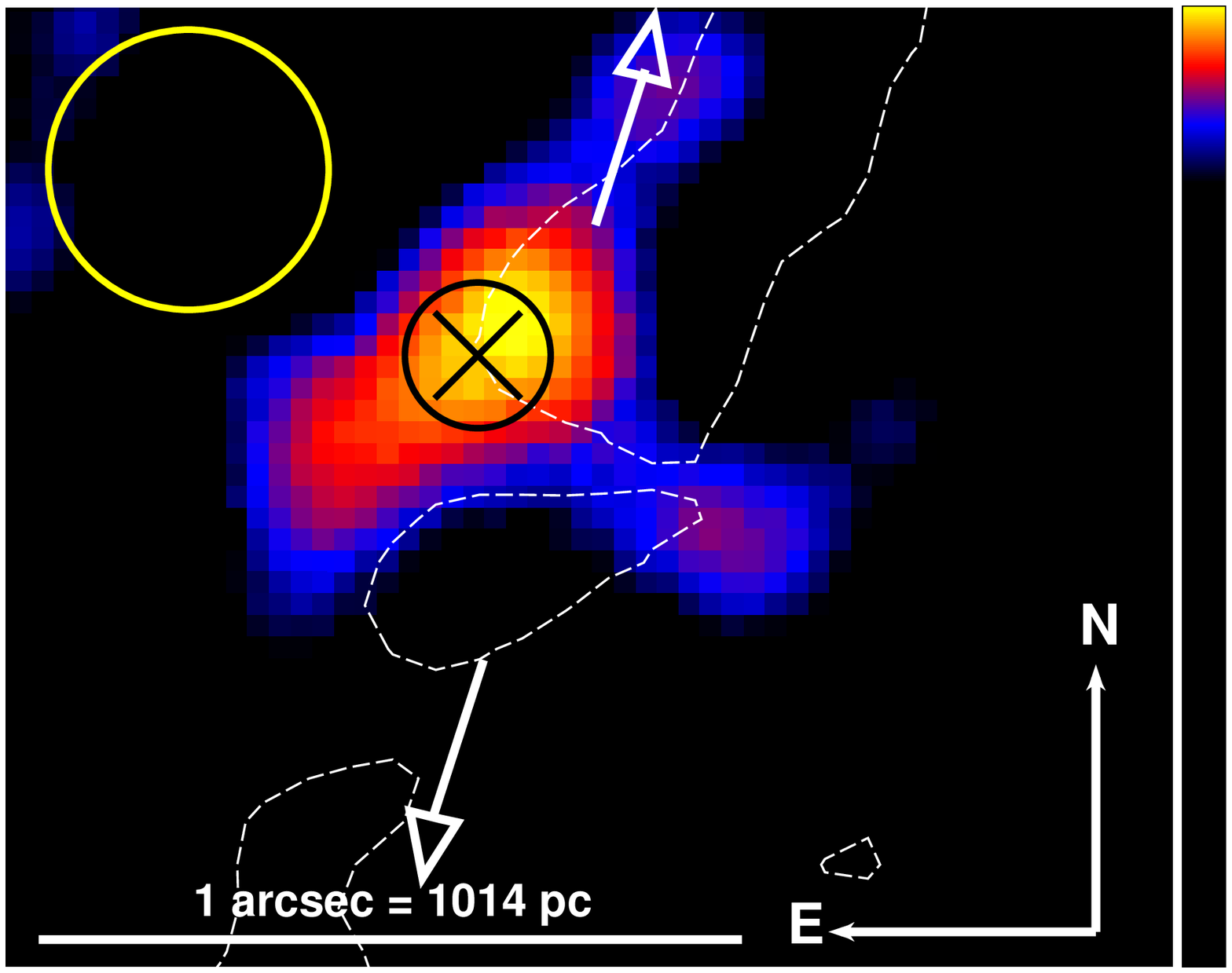}
	\put(55,65){\color{white}\bf 6.3-6.5\,keV}
\end{overpic}%
\begin{overpic}[scale=0.29]{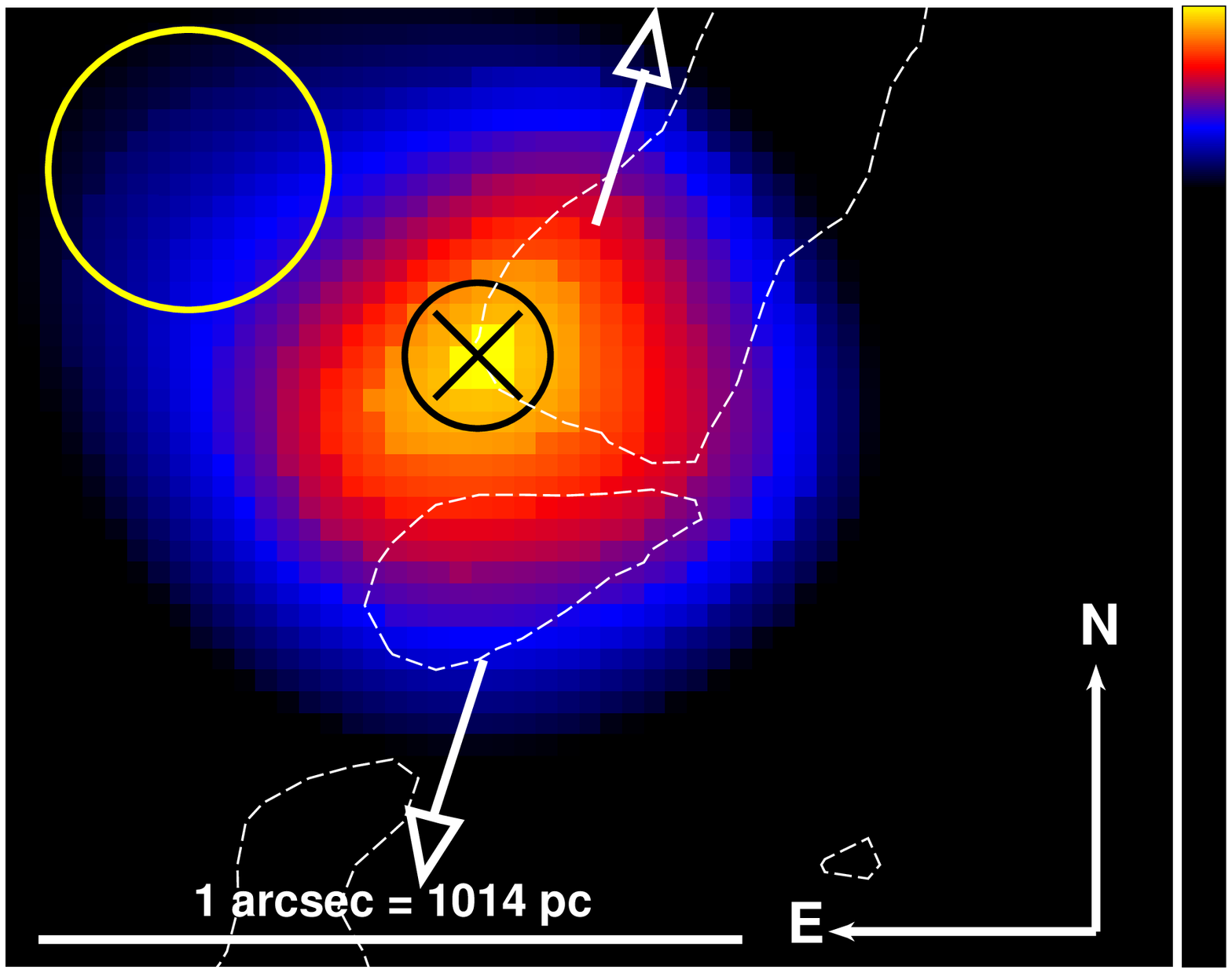}
	\put(55,65){\color{white}\bf 6.3-6.5\,keV}
\end{overpic}%
\includegraphics[width=0.29\textwidth]{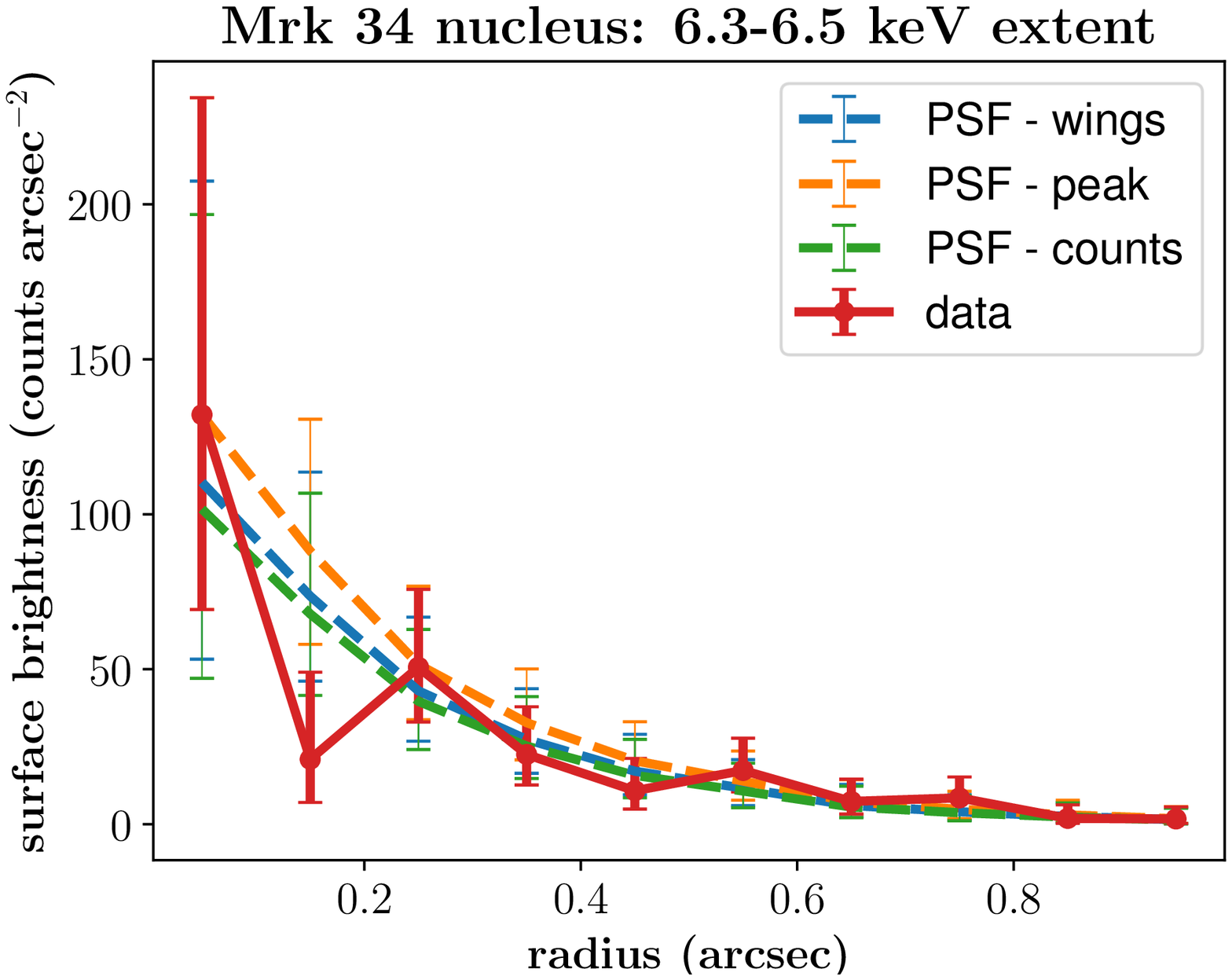}\\
\begin{overpic}[scale=0.29]{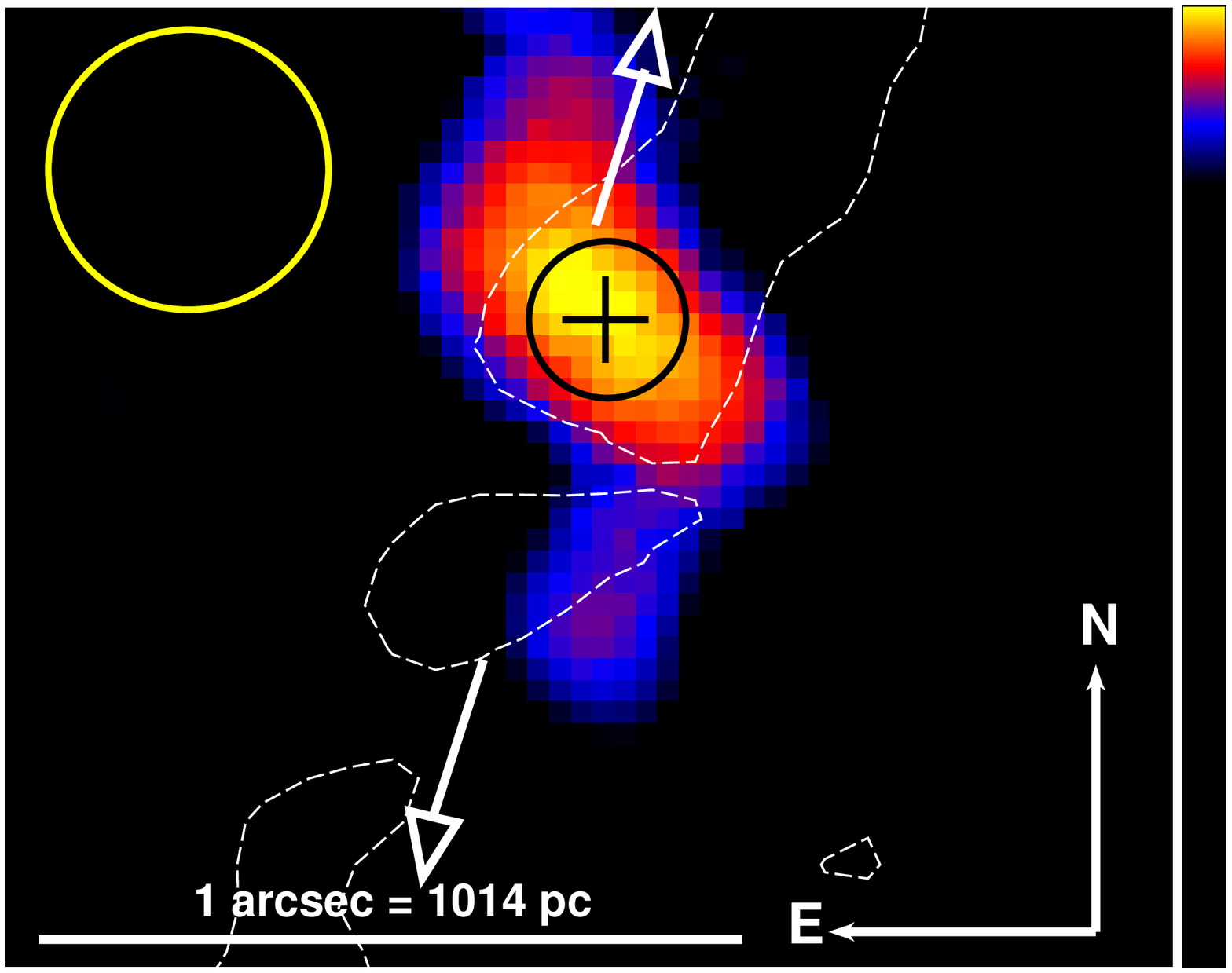}
	\put(55,65){\color{white}\bf 6.5-7.4\,keV}
\end{overpic}%
\begin{overpic}[scale=0.29]{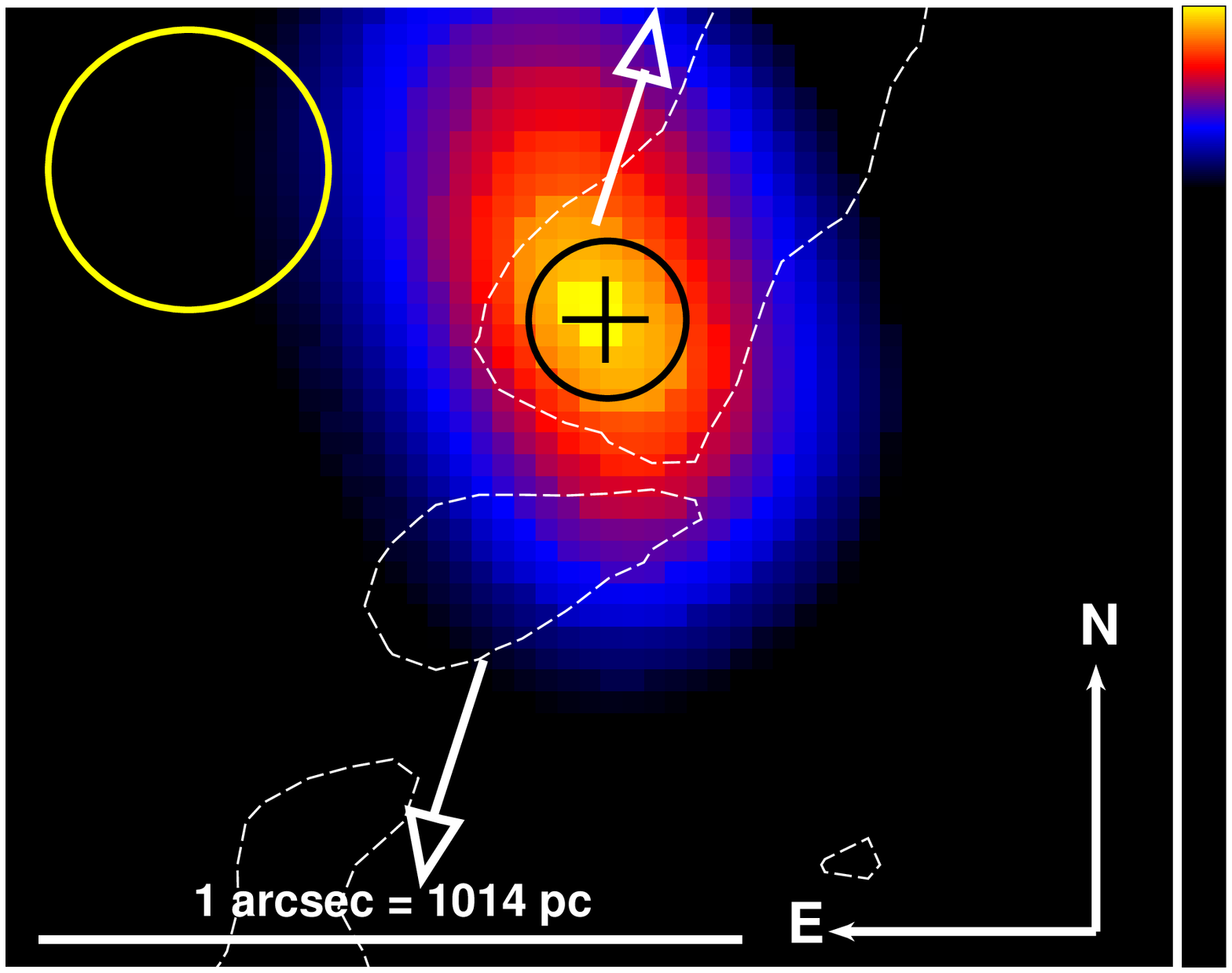}
	\put(55,65){\color{white}\bf 6.5-7.4\,keV}
\end{overpic}%
\includegraphics[width=0.29\textwidth]{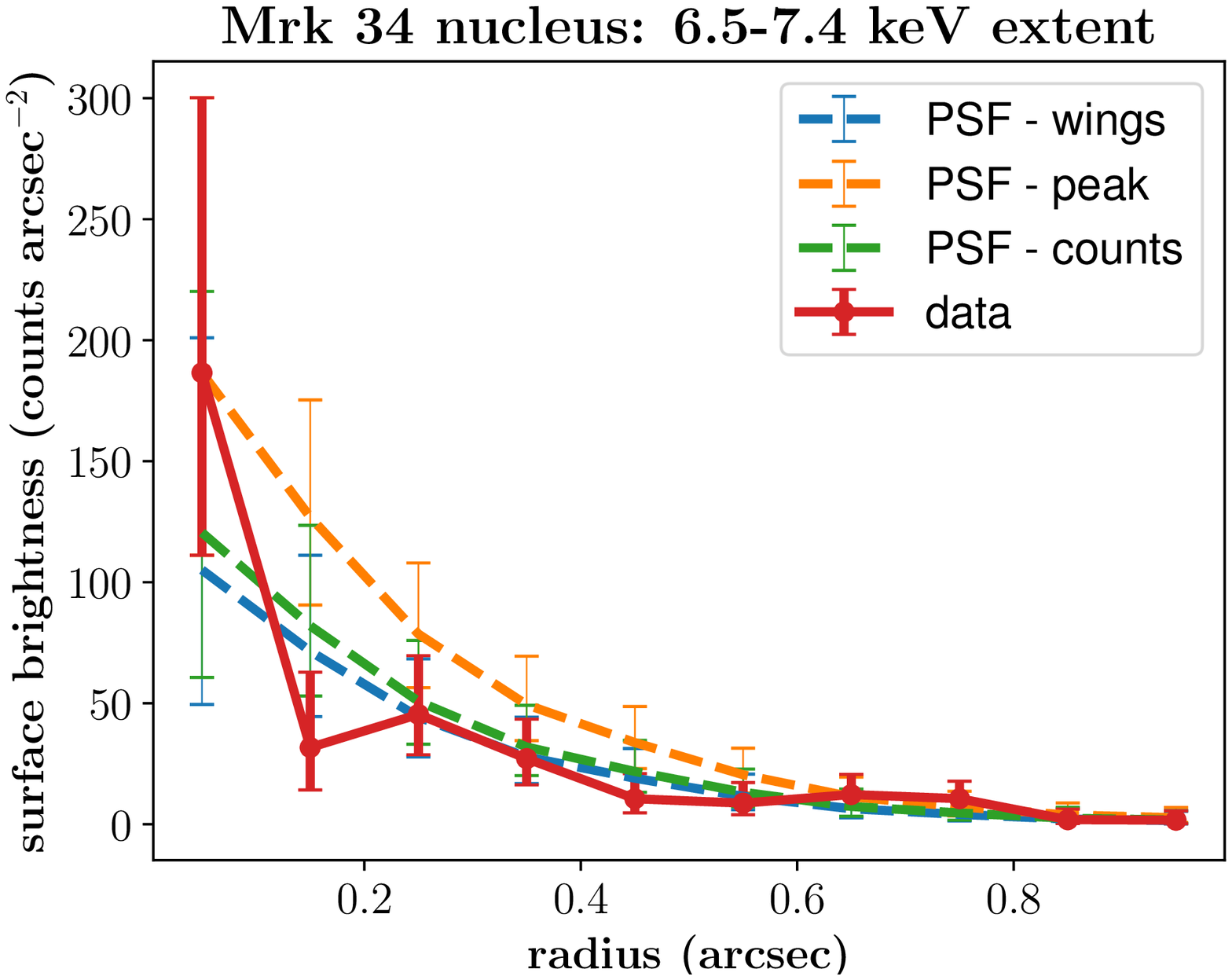}
\caption{{\bf Top to bottom:}  Smoothed images and source extent plots, with rest wavelength bands as labeled.   The 3.2-5.8\,keV and 5.8-6.3\,keV bands show statistically significant deficits in the core relative to the PSF models.  The extended emission evident under different methods of image smoothing is therefore not an artifact of the PSF or photon scatter.  {\bf Left column:}  Images of the Mrk 34 nucleus binned at 1/16th native ACIS pixel scale, with $r=0\farcs215$ Gaussian smoothing applied.  White overlaid contours are from 8.44\,GHz VLA imaging.  Black circles show the centroid positional uncertainty {as determined by {\tt dmstat}}.   The diameter of the yellow solid circle indicates the FWHM of the {\it Chandra} PSF across these bands ($D_{FWHM}\sim0\farcs40$).  White solid arrows indicate the direction and position angle of the fastest-moving optically emitting material according to \cite{Revalski18}.  {\bf Center column:}  As before, but data are adaptively smoothed to preserve features with at least 9 photon counts.  {\bf Right column:}  Radial profiles for band-specific data (solid red) compared to PSF models (dashed) with counts and errors normalized according to median surface brightness for $0\farcs0<r\le0\farcs5$ bins ({\it wings}, blue), {the $0\farcs0<r\le0\farcs1$ bin only} ({\it peak}, orange), and total counts ({\it counts}, green). 
} 
\label{fig:imprep}
\end{figure*}
\clearpage
\makeatletter\onecolumngrid@pop\makeatother

\noindent the NLR and is sufficiently broad that it would be affected if \ion{Fe}{25} contributes significantly.  The residual structure also suggests that a narrow Fe\,K$\alpha$ line should be considered.

{When we assume a single unresolved Gaussian at Fe\,K$\alpha$, residual} excess ``wings" are seen immediately redward (5.8-6.3\,keV, rest) and blueward (6.5-7.0\,keV, rest) of Fe\,K$\alpha$, suggestive of outflows at $v\sim\pm15,000\,\rm{km\,s}^{-1}$ ({assuming an ionization state of \ion{Fe}{17} at most}; Figure \ref{fig:spec}).

Using additional Gaussian components, we estimate the (red, blue) wing to have $[3.0\sigma, 4.1\sigma]$ significance and equivalent width $EW=[0.43\pm0.14\,\rm{keV}, 0.99\pm0.24\,\rm{keV}]$.  The center of the blue wing is at $\sim6.76\pm0.05$\,keV (rest), and is therefore consistent with \ion{Fe}{25} at $v\lesssim2000\,\rm{km\,s}^{-1}$.  {Narrow \ion{Fe}{26} is consistent with a $\sim1\sigma$ detection, and $EW=0.16\pm0.16\,\rm{keV}$.}

For comparison, also in Fig. \ref{fig:spec} we show a fit to a physically motivated model which also incorporates very hard X-ray {constraints} ($>8$\,keV).  

{We also adopt a variant of the \cite{Zhao20} model, which} used a combination of an AGN torus model ({\tt borus02}; \citealt{borus}) and absorbed power laws with high energy cutoffs in a joint fit to this dataset and {\it NuSTAR}.  

{\cite{Zhao20} fit this model using several types of constraints, but we begin with the ``free" fit parameters from \cite{Zhao20} Table 2. These parameters are comparable to other configurations which they consider, which assume [\ion{O}{3}]-derived constraints or $60\degr$ torus inclination.  In order to avoid} model-dependent contamination of the hard continuum model via constraints from complex soft (0.3-3\,keV) extended emission, we excluded the \cite{Zhao20} soft thermal plasma component and froze all other parameters except a normalization constant.  {The spectral shape of the unresolved obscured emission is therefore constrained by the NuSTAR data.  But with $N_H\simgreat10^{25}\,\rm{cm}^{-2}$ from the torus the hard {\it Chandra} continuum effectively reduces to the absorbed power law}.

Excluding the Fe\,K$\alpha$ complex ($\sim5.5-7.0$\,keV observer) from the fit tends to produce excess in the 6.0-6.2\,keV band for the  \cite{Zhao20} model, corresponding to narrow emission from neutral Fe\,K$\alpha$ at 6.4\,keV (rest).  Leaving the {\tt borus02} iron abundance free can reduce this excess, but with unphysically low values for an AGN nucleus ($\sim0.4$).  This suggests that the binning scheme used by \cite{Zhao20} is too coarse {(20 counts\,bin$^{-1}$)} to spectrally resolve the relevant substructure seen here in the Fe\,K$\alpha$ complex and continuum.

Fig. \ref{fig:spec} shows that the excess ``wing" emission observed above the power law is nearly identical compared to this more physically motivated model, and therefore does not appear to be strongly model-dependent.

\subsection{Multi-band Imaging}

\begin{figure}
\vspace{0.1in}
\noindent
\centering
\includegraphics[width=0.47\textwidth]{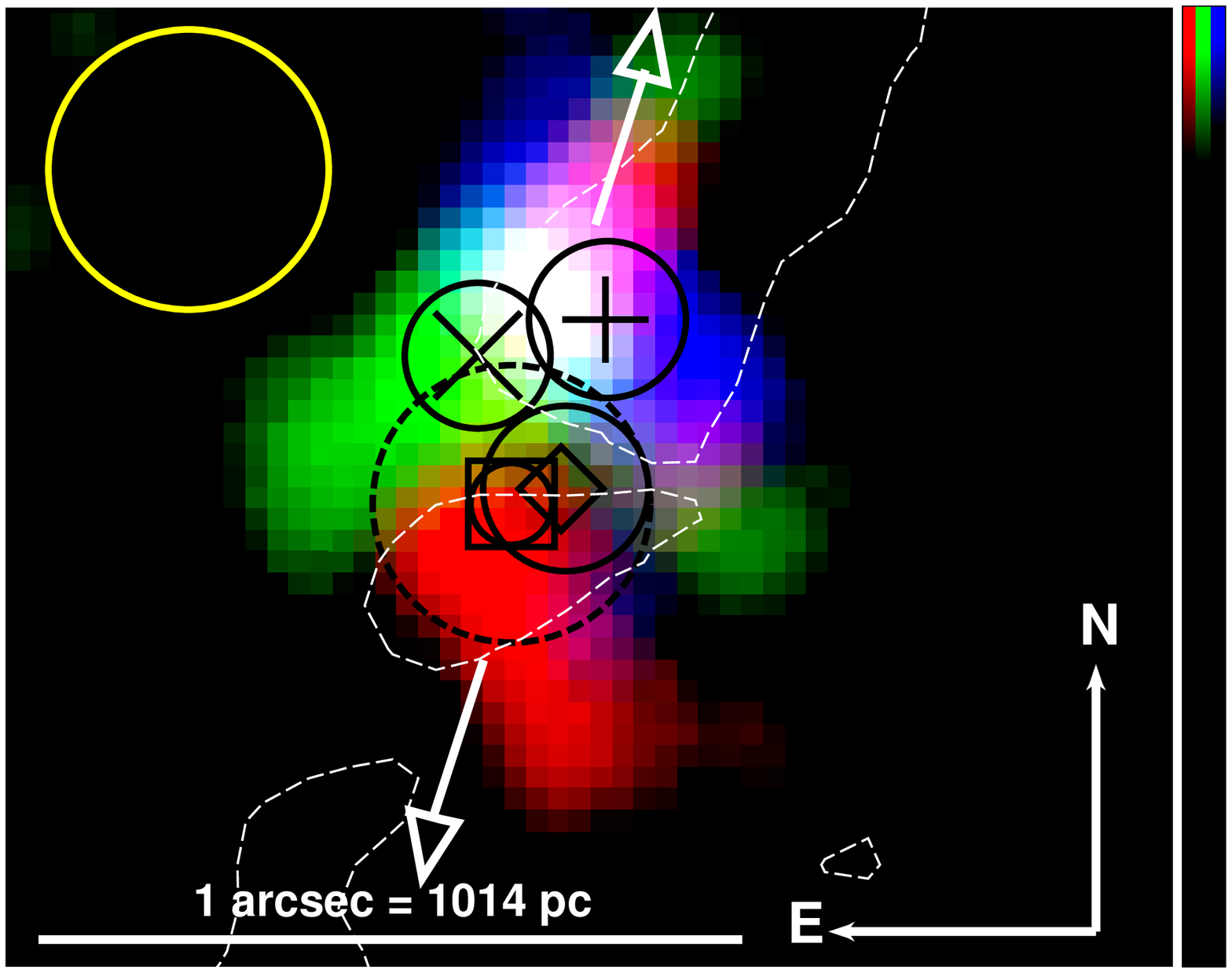}\\
\includegraphics[width=0.47\textwidth]{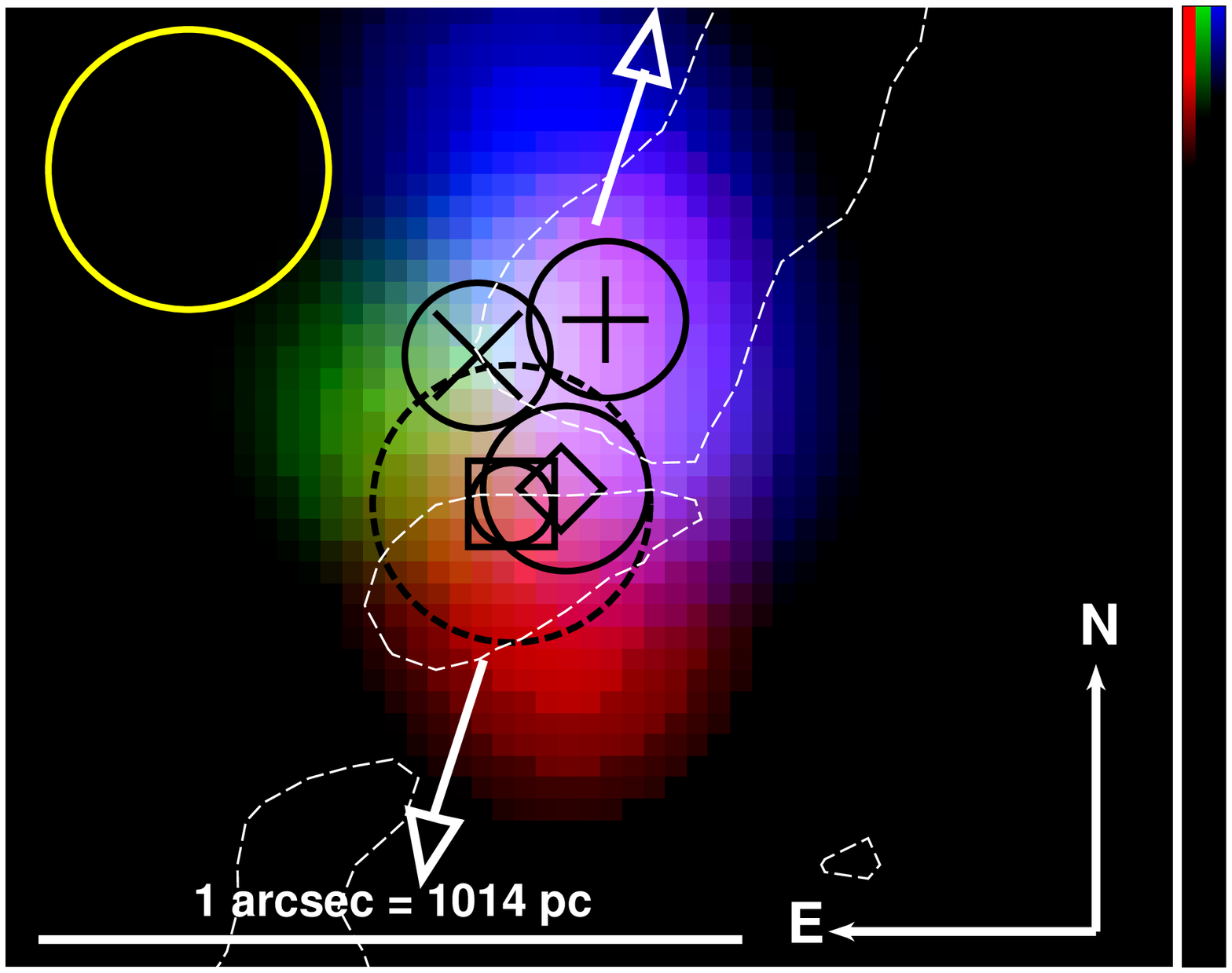}
\caption{{\bf Top:} Multi-color image of the Mrk 34 nucleus binned at 1/16th native ACIS pixel scale, with $r=0\farcs215$ Gaussian smoothing applied.  White overlaid contours are from 8.44\,GHz VLA imaging.  Different image colors indicate the rest 5.8-6.3\,keV band (red), 6.3-6.5\,keV (green) and 6.5-7.4\,keV (blue), as shaded in Fig. \ref{fig:spec}.  Centroid coordinates are indicated with black marks for the 3.2-5.8\,keV rest band (box-circle), 5.8-6.3\,keV (diamond), 6.3-6.5\,keV (`X') and  6.5-7.4\,keV (cross). Black circles show the centroid positional uncertainty.  The diameter of the yellow solid circle indicates the FWHM of the {\it Chandra} PSF across these bands ($D\sim0\farcs40$).  White solid arrows indicate the direction and position angle of the fastest-moving optically emitting material according to \citep{Revalski18}.  {\bf Bottom:}  As above, but data are adaptively smoothed to preserve features with at least 9 photon counts.
} 
\label{fig:img}
\end{figure}

In order to examine whether the possible Fe\,K$\alpha$ wings have spatial extent and structure, we produced images dividing Fe\,K$\alpha$ into three bands (`red wing', 5.8-6.3\,keV rest; 'core', 6.3-6.5\,keV; `blue wing' blue, 6.5-7.4\,keV).  For comparison, we have also included a hard `continuum' band ($3.2-5.8$\,keV rest), which has long been broadly assumed to be point-like in AGN but has recently been observed to have extent in numerous instances \citep[e.g.][]{Maksym17,Ma20,Jones21}.

We binned the photon events at 1/16th native ACIS pixel scale ($0\farcs0625$) and applied Gaussian smoothing with $r=7$ pixels ($0\farcs215$).  We also used {\tt dmimgadapt} to generate adaptively smoothed images that preserve features with $\ge9$ photon counts ($\sim3\sigma$ significance).   VLA radio contours are overlaid to enable comparison to analysis of the Mrk 34 outflows in other work.  We also used a spectrum of the nucleus, {\tt marx} \citep{MARX} and SAOTrace\footnote{\url{https://cxc.harvard.edu/ciao/PSFs/chart2/runchart.html}} to simulate an event file of the ACIS point spread function with $\sim2600$ 5.5-7.0\,keV counts at $r<2\arcsec$ ($\sim10\times$ observed).  

Figure \ref{fig:imprep} shows the single band results of this smoothing (column \#1 for Gaussian smoothing, column \#2 for adaptive smoothing).  Figure \ref{fig:imprep} column \#3 shows radial profiles of the Mrk 34 nucleus for $r\le1\farcs0$, binned at $\Delta r=0\farcs1$ intervals.  The band-specific centroid for {each unsmoothed photon image is set to $r=0$ and compared against} the band-specific PSF simulation.  The PSF is shown with counts and errors according to three different normalizations: median surface brightness for the $0\farcs0<r\le0\farcs5$ bins, the $r\le0\farcs1$ bin only, and total counts.  

The hard continuum (3.2-5.8\,keV, 114 counts for $r<2\farcs0$) shows two features roughly NW and SE of the band centroid, separated by $\sim1$ PSF FWHM.  {The radial profile of this band shows a central $r\le0\farcs1$ bin with} a significant deficit {when normalized to match the PSF wings}, and its $\chi^2$ test is inconsistent with the PSF at $3.0\sigma$.

The red wing ({5.8-6.3\,keV}, 41 counts for $r<2\farcs0$) also shows significant resolved structure.  It also has a significant radial profile deficit at $r\le0\farcs1$.  The $\chi^2$ radial profile alone rejects the PSF model at only 90\% significance, but there is also significant azimuthal structure with a N-S orientation.  When $0\farcs1<r<1\farcs0$ data are divided into four equal N, S, E and W bins, the PSF model is rejected at $3.1\sigma$.

PSF comparisons fail to show extent within the blue wing (6.5-7.4\,keV rest, 42 counts, $p\sim0.5$) and core Fe\,K$\alpha$ band (6.3-6.5\,keV, 46 counts, $p\sim0.2$).  

For cross-band comparison, we produced an image (Fig. \ref{fig:img}) which overlays the Fe\,K$\alpha$ bands (red, 5.8-6.3\,keV rest; green 'core', 6.3-6.5\,keV; blue, 6.5-7.4\,keV), as well as centroid positions for the Fe\,K$\alpha$ bands and the hard continuum.

{Positional uncertainties are determined via {\tt dmstat} and are typically $\sim0\farcs1$}.  The centroid positions of the red, core and blue wings of the Fe\,K$\alpha$ complex all vary by $\simgreat2\sigma$ from each other.  The hard continuum centroid is consistent with the other bands, but its large positional uncertainty is attributable to observed NW-SE substructure larger than the PSF uncertainty in the Fe\,K$\alpha$ complex bands.  There is therefore spatial structure not only within bands (in the hard continuum and red wing) but via intra-band comparison across all four bands (including also the Fe\,K$\alpha$ core) at the limits of the {\it Chandra} PSF.

\begin{figure}
\vspace{0.1in}
\noindent
\centering
\includegraphics[width=0.47\textwidth]{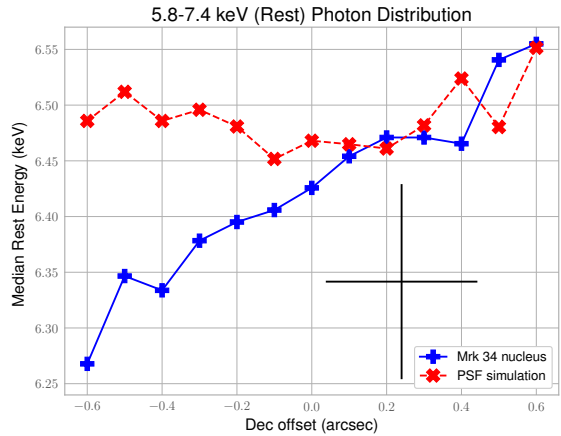}
\caption{Plot of the median energy of a $1\arcsec\times0\farcs5$ box in ($\alpha$,$\delta$) for the 5.8-7.4\,keV rest band.  Blue `+' marks correspond to values from the observations of Mrk\,34 with {\it Chandra}.  A {\it Chandra} PSF simulation (red `X' marks) has no such linear energy-dependent trends, but rather an expected small increase with radius.  The simulation has a median energy offset attributable to slightly different conditions in the PSF seed spectrum, which uses a smaller extraction radius ($r=1\arcsec$ vs $r=2\arcsec$) and applies a point source correction.  The representative uncertainty (black cross) indicates the energy and spatial resolutions ($\pm$FWHM/2, with redshift correction for photon energy).}
\label{fig:medE}
\end{figure}

In order to further quantify this observed energy-dependent spatial structure in terms of observed N-S trends (and for comparison against outflow structures observed in the radio and optical, e.g. \citealt{Revalski18}), we calculated the median photon energy as a function of $\delta$ offset in the 5.8-7.4\,keV rest band for a series of $1\arcsec\times0\farcs5$ boxes in ($\alpha,\delta$), each offset by $0\farcs1$ in $\delta$.  The result is shown in Fig. \ref{fig:medE}, along with a similar energy profile for the PSF simulation.  The median energy of the Mrk 34 nucleus shifts almost monotonically by 288\,eV ($\sim13,000\,\rm{km\,s}^{-1}$) over $1\farcs2$ ($\sim1.2$\,kpc).  This exceeds both the ACIS-S energy FWHM of $\sim160$\,eV at 5.9\,keV, and the in-band simulated PSF FWHM of $\sim0\farcs4$).  The bins' median energies for the PSF simulation span $\simless100$\,eV.  This span is greatest between the core and larger radii, and likely dominated by the tendency of the PSF to spread at larger energies\footnote{\url{https://cxc.harvard.edu/proposer/POG/}}.

\section{Discussion}

{\it Chandra} ACIS imaging spectroscopy of the Mrk 34 nucleus shows evidence for spatially resolved ($D\simgreat\,500$\,pc) excess emission in the red and blue wings of Fe K$\alpha$, which corresponds to projected velocities of $\sim15,000\,\rm{km\,s}^{-1}$, and is robust against both physically motivated broad-band models \citep{Zhao20} and a simple power law continuum.  Given the Compton-thick column density inferred by \cite{Zhao20}, this spectral structure is unlikely to be attributable to the accretion disk, which would be highly obscured.  Rather, the Fe\,K$\alpha$ complex emission should be emitted from the torus which would not be extended, $r\simgreat\rm{few}\times0.6$pc from the bolometric luminosity \citep{TF21} and the dust sublimation radius \citep{Nenkova08a,Nenkova08b}, or from beyond the torus.

The hard continuum and Fe\,K$\alpha$ complex show clear evidence for spatial extent, both within bands (in the hard continuum and red wing) and across bands (including also the Fe\,K$\alpha$ core), ruling out a torus origin.  We observe a spatially resolved trend of increasing median photon energy from south to north within the Fe\,K$\alpha$ complex.  This trend and the observed N-S structure within the red wing suggest	 a southern structure dominated by red wing photons, as well as a northern structure which is more spectroscopically complex, containing the blue wing and possibly contributions from other parts of the Fe\,K$\alpha$ complex.  This structure may be related to the dual peak structure seen in the hard continuum.

The representative scale of the multiband N-S structure between the centroid of the blue wing and the southernmost resolved features of either the hard continuum or the red wing is $D\simgreat0.4\arcsec$, which corresponds to $r\simgreat200$\,pc.  We can assume bilateral symmetry from a central nucleus and adopt the same bi-cone geometry as found for the [\ion{O}{3}] outflow.  The projected morphological orientation of the Fe\,K$\alpha$ complex is consistent with structure described in other bands \citep{Falcke98,Fischer13,Revalski18,TF21}.

%The observed velocities of +/- 7500 km/s then correspond to true space velocities of +/- 21,000 km/s. Compared to the maximum deprojected [OIII] velocities of ~2000 km/s, the X-ray outflows would then carry ~100 X more kinetic energy per unit mass.

\subsection{Low-ionization Fluorescence}

\begin{figure}
\vspace{0.1in}
\noindent
\centering
\includegraphics[width=0.47\textwidth]{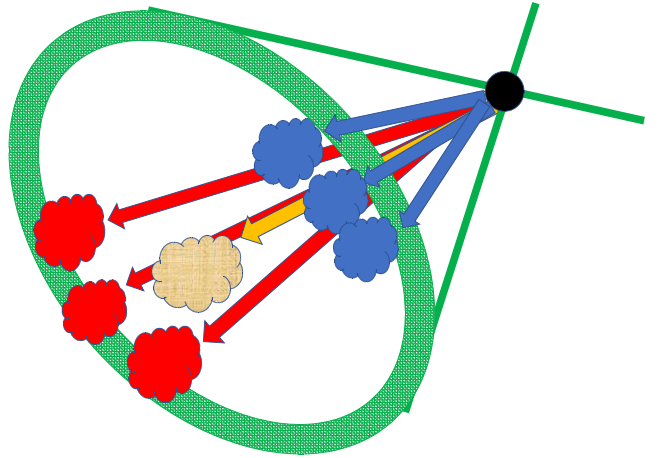}
\caption{{Top: Cartoon based upon \cite{Fischer13}, one of the possible models we consider (see also Fig. \ref{fig:conemod}). The observer is above the image and the SE cone is illustrated.  Green indicates the bounds of the hollow biconical outflow seen in moderate-ionization optical ENLR gas (e.g. [\ion{O}{3}]) which may be dominated by wind-ISM interactions.  Arrows indicate X-ray outflow velocity projected in the plane of the sky.  Brown clouds have zero line-of-sight velocity relative to the host.  Redshifted clouds (red) recede from the observer due to the geometry and inclination of the cone.  Due to the inclination of the cone, their velocity projected to the observer is small relative to the waveband under consideration, allowing them to be detected.  Blueshifted clouds (blue) approach the observer.  Their velocity projected to the observer is large relative to the waveband and the clouds are undetected.  In the NW cone the situation is similar but instead the blueshifted clouds remain in-band.  The X-ray outflows are interior to the conical bounds and retain a higher velocity than the optically emitting gas.
}
}
\label{fig:cartoon}
\end{figure}
\begin{figure}
\vspace{0.1in}
\noindent
\centering
\includegraphics[width=0.47\textwidth]{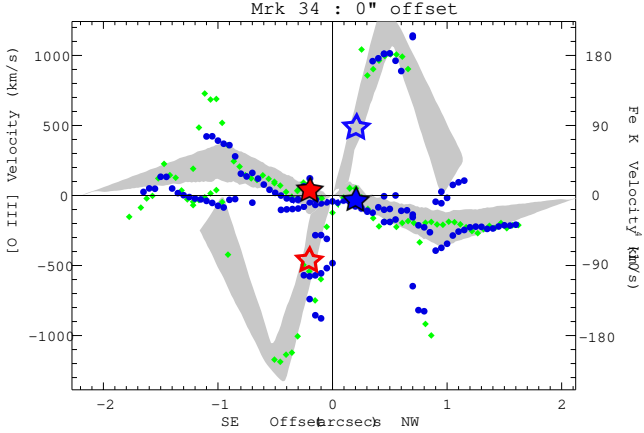}
\caption{Fig. 13 from \cite{Fischer13}, modified to illustrate the effects of an inclined hollow bi-cone geometry on projected velocities {(see Fig. \ref{fig:cartoon})}.  Green and blue circles represent [\ion{O}{3}] velocities as observed by {\it HST}, with velocity scale indicated on the {\bf left side Y-axis}.  Grey shading represents the projected  [\ion{O}{3}] velocity structure of the bi-cone model.  Red and blue filled stars indicate projected velocities measured by {\it Chandra} for the red and blue wings of Fe\,K$\alpha$, with velocity scale indicated on the {\bf right side Y-axis}.  Red and blue empty stars indicate the unmeasured component corresponding to gas on the opposite exterior side of the hollow cone, with larger and opposite projected velocities.}
\label{fig:conemod}
\end{figure}

\begin{figure}
\vspace{0.1in}
\noindent
\centering
\includegraphics[width=0.47\textwidth]{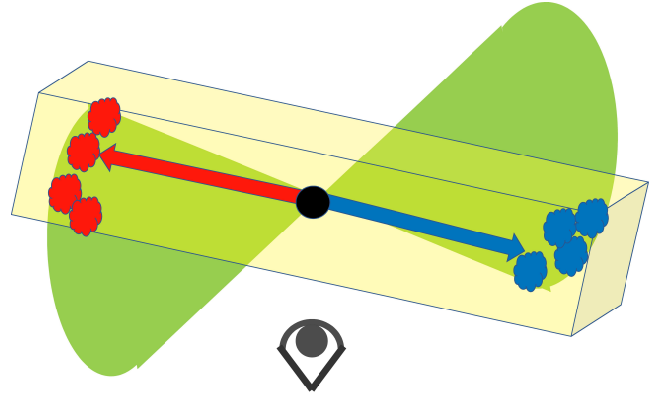}
\caption{{Top: Cartoon based upon \cite{Revalski18,Revalski19}, an alternative to the \cite{Fischer13} model. The observer is at bottom looking up (dark grey).  Green indicates the hollow biconical outflow seen in moderate-ionization optical ENLR gas (e.g. [\ion{O}{3}]) which may be dominated by wind-ISM interactions. The yellow box is the plane of the galaxy, which is a large reservoir of gas for wind-ISM interactions.  Red and blue indicate X-ray outflows seen in the red and blue wings of Fe\,K$\alpha$ which are redshifted and blueshifted accordingly.  The X-ray outflows are interior to the conical bounds and retain a higher velocity than the optically emitting gas.
}
}
\label{fig:cartoon2}
\end{figure}

Although some of the blue wing may arise from thermal outflows, fluorescence provides a simpler explanation than thermal outflows for various reasons.  First, the large EW of the red and blue wings ($\sim[0.43,0.99]$\,keV) suggests reflection by the K$\alpha$ transition of nearly neutral Fe species, such as might be found in dust, molecular clouds, or a moderately ionized X-ray wind (K$\alpha$ transitions from $\simless$ \ion{Fe}{21} are confused with \ion{Fe}{1} by the ACIS-S CCD).

A spatially resolved outflow could explain both red and blue wings via bilateral symmetry similar to that seen from low-velocity outflows in optical and radio emission.  If we assume the \cite{Fischer13} bi-conical outflow model (SE, axial inclination $i=25\degr$ towards the observer from the plane of the sky) and the blue-red N-S orientation of the Fe K$\alpha$ wings, then the red and blue wings have the opposite projected velocities expected from a simple bulk outflow near the bi-cone axis.  {Figs. \ref{fig:cartoon} and \ref{fig:conemod} illustrate the velocity structure of this model in comparison with results from our data.}

The velocity splitting produced by [\ion{O}{3}] in the \cite{Fischer13} outflow model suggests that the X-ray emitting gas might also demonstrate velocity splitting due to a hollow conical structure.  If so, the blue-red N-S orientation of the Fe K$\alpha$ wings requires them to be produced at an angle in the bi-cone which has a small inclination from plane of the sky $i_{\rm Fe}$ and small projected velocities, such that $\theta_{min}-i\le i_{\rm Fe}\le\theta_{max}-i$ (where $\theta_{min}=30\degr$ is the inner opening angle {and $\theta_{max}=40\degr$ is the outer opening angle)}.

This model implies a negligible spatial deprojection, $\simless4\%$ difference.  The observed spatial extent of the Fe\,K$\alpha$ structure, $r\simgreat200$\,pc, is $\simgreat100\times$ larger than the torus.  The $r\simless200$\,pc region is associated with rapid radial increases in optical kinetic energy, kinetic power momentum and momentum outflow, and the X-ray features we describe are interior to the associated turnovers in these quantities at $r\sim500$\,pc \citep{Revalski18,Revalski19}. 

{Since $3.9\simless({\rm sin}\,i_{\rm Fe})^{-1}\simless11.5$, the median photon energy shift} in Fig. \ref{fig:medE} ($\sim150$\,eV) from the central bin corresponds to deprojected $0.09c \simless |v| \simless 0.26c$.  The red wing model excess is fit as a Gaussian at $\sim6.08$\,keV ($0.2c \simless |v| \simless 0.5c$ deprojected).  The velocity shift for the blue wing would have a similar value.  For comparison, the fastest optically emitting material in Mrk 34 has a deprojected velocity of $v\sim2000\,\rm{km\,s}^{-1}$, with the same [N, S] orientation for [blue, red] shifts.

Complementary counterpart features produced on the opposite side of the Mrk 34 cone are likely to be difficult to observe with {\it Chandra} given the corresponding projection angles.  Redshifted Fe K$\alpha$ gas in the NW cone may suffer from obscuration by host ISM \citep[e.g. a likely E-W nuclear dust feature seen by {\it HST}][]{Falcke98}, relativistic de-beaming and greater low-energy continuum from spatially confused emission.  The energy response of {\it Chandra} ACIS drops more rapidly above $\sim7$\,keV than any gain from 
Blueshifted Fe K$\alpha$ gas in the SE cone would fall, implying a $\simless1\sigma$ detection for a red wing counterpart projected at a blueshift of $v\simgreat0.2c$.  By comparison, the $0.15\simless E/{6.39\,{\rm keV}}-1 \simless0.50$ energy band (systemic; 7.0-9.1\,keV observer) contains [4, 8] photons for $r\le[0\farcs5,1\farcs0]$, including continuum.  This band's centroid is $\sim0\farcs4$ south of the blue wing centroid, and consistent with the southern feature in the red wing.

Unlike \cite{Fischer13}, \cite{Revalski18,Revalski19} assume that the [\ion{O}{3}] gas in Mrk 34 is confined to the plane of the galaxy {(Fig. \ref{fig:cartoon2})}.  In this model, the velocity deprojection factor would only be $\sim1.56$, implying deprojected $v\sim0.08c$.  This interpretation is complicated by the lack of measured [\ion{O}{3}] outflow components with the appropriate blue/red shift direction at $r\simless0\farcs5$ {but unlike  \cite{Fischer13} does not require explaining the nondetection of an opposite-shift counterpart in the same cone}.

In either model, the deprojected velocities would be solidly in the regime of ultrafast outflows which in X-rays have normally been seen (unlike Compton-thick Mrk 34) as absorption at $0.03c\simless v \simless 0.3c$ in an otherwise weakly obscured continuum, and commonly attributed to \ion{Fe}{25} or \ion{Fe}{26}, \citealt{Tombesi10,Tombesi14,Gofford13,Chartas21} (but see also a P-Cygni-like profile in \citealt{Nardini15}).  The velocity structure of Mrk 34 suggests bulk motion $\simgreat12-30\times$ faster than the optical gas.  With kinetic energy $KE_{X-ray}\simgreat140-800\times(M_{X-ray}/M_{optical})$, such gas could produce the kinetic power in the X-ray winds described by \cite{TF21} with $M_{X-ray}/M_{optical}\simless0.06-0.34$.  Such winds might easily carry the bulk of the outflow kinetic energy and drive the acceleration of [\ion{O}{3}] gas.

Since strong Fe\,K$\alpha$ is likely to require a large column density ($log N_H\simgreat23$), it is likely to be associated with dusty molecular clouds.  Such dense, low-ionization gas might then be converted into an [\ion{O}{3}] wind expanding at larger radii, as per \cite{TF21}.  As per the galaxy-scale outflows in IC 5063, such clouds may be launched from the ISM via ablation at small radii \citep{Maksym20,Maksym21,Travascio21} and entrained, or near the plane of the galaxy.  Alternately, high-velocity dust may originate in the AGN wind itself, as outflowing clouds condense to form molecular structures as they cool \citep{Elvis02}.

\subsection{Spatially Resolved Obscuration}

Local column overdensities may contribute to some of the observed spatial structure.  {\it HST} observations \citep{Falcke98} show evidence for dust lanes which cut E-W across the nucleus.  Such dust lanes could, for certain column densities, suppress the relatively soft continuum emission and red wing, e.g. at the spatial location of the blue wing.  {The blue wing is spatially coincident with the brightest radio hot spot, which may indicate the nucleus (such that deprojection may not be appropriate for the blue wing), but the absolute and relative astrometric uncertainty do not require this.  If the blue wing is coincident with the nucleus, then the inferred radial extent is greater for the southern outflow (which is receding relative to the observer) but implies a larger radial extent for that outflow and implies a lack of a northern biconical counterpart.}

\subsection{Thermal Outflows}

The blue wing energy range is consistent highly ionized \ion{Fe}{25} at $v\simless2,000\,\rm{km\,s}^{-1}$, which corresponds to $v\simless3,000-22,000\,\rm{km\,s}^{-1}$ depending upon the deprojection model.   Others have observed extended \ion{Fe}{25} in association with strong AGN outflow-ISM shocks (e.g. NGC 6240, \citealt{Wang14}; NGC 4945, \citealt{Marinucci17}; and IC 5063, \citealt{Travascio21}).  The proximity of the blue wing to the peak 8.44 GHz may point to a similar scenario in Mrk 34.  

The spatial extent of the observed structure is consistent with that of optically emitting gas in a region where the kinetic power of the optically emitting gas rises steadily with increasing radius \citep{Revalski18,Revalski19}.  {This could be true whether the blue wing were associated with the nucleus or offset from it. If the emission does originate} from the continuum or highly-ionized \ion{Fe}{25} then it may point to shocks in the hot component of a multi-phase medium. Simulations by e.g. \citealt{Mukherjee18} indicate the presence for $kT\simgreat10^8$\,K gas in IC 5063.   \ion{Fe}{25} in Mrk 34 suggests $kT\sim10^{7.5}$\,K, {and for \ion{Fe}{26} to be negligible} $kT$ would be narrowly constrained.

One major difficulty in this scenario includes the large wing EW.  In addition, the red wing emission is particularly difficult to explain via \ion{Fe}{25}, requiring deprojected $v\simgreat0.4c$.  If associated with bipolar outflows, such a feature may leave similarly blueshifted features at other energies (possibly via other species), which we have not been able to identify.

{In principle, ``Back-streaming" from jet-ISM interactions could cause a combination of low-velocity and high-velocity features.  For example, \cite{Das05} claim such effects for faint, high velocity [\ion{O}{3}] clouds in NGC 4151 as the result of shocks.  If some of the red and blueshifted iron line emission is co-located (such as with the northern component of the spatially bifurcated red wing, which is spatially consistent with the blue wing and the brightest radio hotspot) then \ion{Fe}{25} and possibly \ion{Fe}{26} could be formed at a shock, while redshifted emission could be part of the backflow.
} 

\subsection{An Obscured Relativistic Jet}
Bremsstrahlung from a relativistic jet at the nucleus might mimic an excess spanning the red and blue wings when obscured by Compton-thick material.  Such an excess might be variable on observable timescales, and would appear as continuum emission at microcalorimeter (e.g. {\it XRISM} or {\it Athena}) resolution.  Doppler effects from sufficiently relativistic extended outflows (either a wind or jet) might affect assumptions of spatial symmetry.

\section{Conclusion}

The spatially and spectrally extended X-ray emission associated with the Fe\,K$\alpha$ complex of Mrk 34 is likely associated with the well-known powerful kpc-scale outflows which have been studied in optical and radio emission.  This X-ray emission is likely due to redshifted and blueshifted Fe\,K$\alpha$ fluorescence and may be associated with molecular clouds at the brightest radio knots, near the base of the outflow.  Some of this emission may be associated with shock emission from a hot collisionally ionized plasma that also produces highly ionized  \ion{Fe}{25}, as has been observed in IC 5063 \citep{Travascio21}.  

In any event, the source of the emission is likely to be expanding from the nucleus at a velocity a factor of $\simgreat12-28$ greater than outflows observed in optical spectroscopy.  Regardless of the specific origin of the part of emission, then, it is likely to be a multi-phase medium with fast-moving hot gas which may indicate a critical link between sub-pc scale ultra-fast outflows previously seen in absorption, as well as $\sim$kpc-scale feedback mechanisms.  {Using {\it Chandra} observations of NGC 5728, \cite{TF23a,TF23b} have recently demonstrated a new example of emission attributable to Fe\,K$\alpha$ emission which is spatially and spectrally extended, and suggestive of ultrafast outflows on $\simless200$\,pc scales.  Although several aspects of the specific physical interpretation remain uncertain, their work demonstrates larger spatial extent, superior photon statistics, and confirms the existence of systems with properties similar to those we attribute to this novel Fe\,K$\alpha$ complex in Mrk 34.}  Such interactions between the hot outflows and dense molecular clouds near the nucleus may form the base of the expanding [\ion{O}{3}] and X-ray winds {via similar mechanisms to those} described by \cite{TF21}.  

New observations are necessary to better describe the origins of this extended emission.  Deeper imaging spectroscopy with {\it Chandra} or a high-resolution implementation of {\it AXIS} would provide better statistical constraints on these models, and microcalorimeter spectroscopy from {\it XRISM} or {\it Athena} would spectrally resolve composite nuclear models.  {\it NuSTAR} may constrain the presence of a line-of-sight component for the bi-conical outflow.  {\it Lynx} is necessary to unambiguously associate reliable emission models with the spatial features described here.  Imaging spectroscopy with {\it JWST} and {\it ngVLA} is necessary to investigate the roles of outflowing molecular gas and moderately ionized winds, both of which are likely present, regardless of the specific iron excitation mechanisms.

\begin{acknowledgments}

We thank the referee for a very careful review and comments which greatly improved the quality of the manuscript.  We thank Margarita Karovska and Xirui Zhao for helpful discussions.  WPM acknowledges support by Chandra grants GO1-22088X and GO8-19096X.  

\end{acknowledgments}


\vspace{10mm}
\facilities{CXO(ACIS)}


\software{CIAO,  \citep{CIAO}
          XSPEC \citep{XSPEC}, 
         ds9 \citep{ds9},
         MARX \citep{MARX}
          }





\end{document}